\documentclass{article}

     \usepackage[final]{neurips_2019}

\usepackage[utf8]{inputenc} 
\usepackage[T1]{fontenc}    
\usepackage{hyperref}       
\usepackage{url}            
\usepackage{booktabs}       
\usepackage{amsfonts}       
\usepackage{nicefrac}       
\usepackage{microtype}      
\usepackage{graphicx}
\title{Prediction of optical spectra of coarse-grained polymers as a sequence generation problem: the {\it{Recurrent Neural Networks}} solution.
}

\author{%
Lena Simine \thanks{previous affiliation: Department of Chemistry, Rice University} \\
  Department of Chemistry\\
  McGill University\\
  Montreal, QC, Canada \\
  \texttt{lena.simine@mcgill.ca} \\
   \And
   Thomas C. Allen \\
   Department of Chemistry\\
   Rice University \\
   Houston, TX, USA \\
   \texttt{tca1@rice.edu} \\
   \AND
   Peter J. Rossky \\
   Department of Chemistry\\
   Rice University \\
   Houston, TX, USA \\
   \texttt{peter.rossky@rice.edu} \\
}

\begin{document}

\maketitle

\begin{abstract}
  Coarse-grained simulations of conjugated polymers have become a popular way of investigating the device physics of organic photovoltaics. While UV-Vis spectroscopy remains one of key experimental methods for the interrogation of these devices, a rigorous bridge between coarse-grained simulations and spectroscopy has never been established. Here we address this challenge by developing a method that predicts spectra of conjugated polymers directly from coarse-grained representations while avoiding ad-hoc procedures such as back-mapping from coarse-grained to atomistic representations followed by computing the spectra using standard quantum chemistry methods. Our approach is based on a generative deep learning model: the long-short-term memory recurrent neural network (LSTM-RNN) and it is suggested by the apparent similarity between natural languages and the mathematical structure of the perturbative expansions of the excited state energies due to small fluctuations of the polymer conformation. We use this model to demonstrate a dangerous discrepancy between the spectra obtained in the coarse-grained representation and after the back-mapping. This indicates that standard protocols may require additional fine-tuning in order to become reliable and that our model presents a novel tool uniquely suited for improving the back-mapping protocols and for including spectral data in the development of coarse-grained potentials.
\end{abstract}

\section{Introduction}

Coarse-grained (CG) simulations have become a popular way of investigating the device physics of organic photovoltaics. For a recent review see Ref.(1) and references therein. Structural coarse-graining reduces the computational load and increases the time-step thereby helping approach device-relevant time- and length-scales(2, 3). Clearly, coarse-grained simulations have the potential to generate insights into the sought-after structure-function relationship in the active layer of organic solar cells. In order to realize this potential, however, a relationship between the coarse-grained structures and their electronic properties needs to be established. Currently, this is done through back-mapping of the coarse-grained structures into atomistic representations which serve as inputs for quantum chemistry methods. Naturally, a rigorous back-mapping or fine-graining procedure cannot be derived rigorously and instead it is developed in an ad-hoc manner using physically motivated but, in general, uncontrolled approximations. In order to learn about the electronic properties of large aggregates of conjugated polymers from coarse-grained simulations a new methodology which would bypass the back-mapping step is required. Here, we make the first step in this direction by developing a method for predicting UV-Vis absorption spectra directly from coarse grained representations of conjugated polymers using machine learning. See Fig 1A for a summary schematic. This method is based on a deep learning model used for modeling of sequential data such as natural language processing - the Long-Short-Term-Memory Recurrent Neural Net (LSTM-RNN)(4, 5). 

The idea of drawing an analogy between the problem of predicting absorption spectra from incomplete structural input and the problem of language generation is inspired by perturbation theory. The mathematical structure of the perturbative corrections to the spectra of quantum many-electron Hamiltonians due to small fluctuations of a small subset of model parameters is a series of products of these parameters with varying lengths and compositions, much like the words in the English language can be viewed as ‘products’ or sequences of letters varying in length and composition. 

\begin{figure}
  \centering
  \fbox{\rule[-.5cm]{0cm}{4cm} \includegraphics[width=\textwidth]{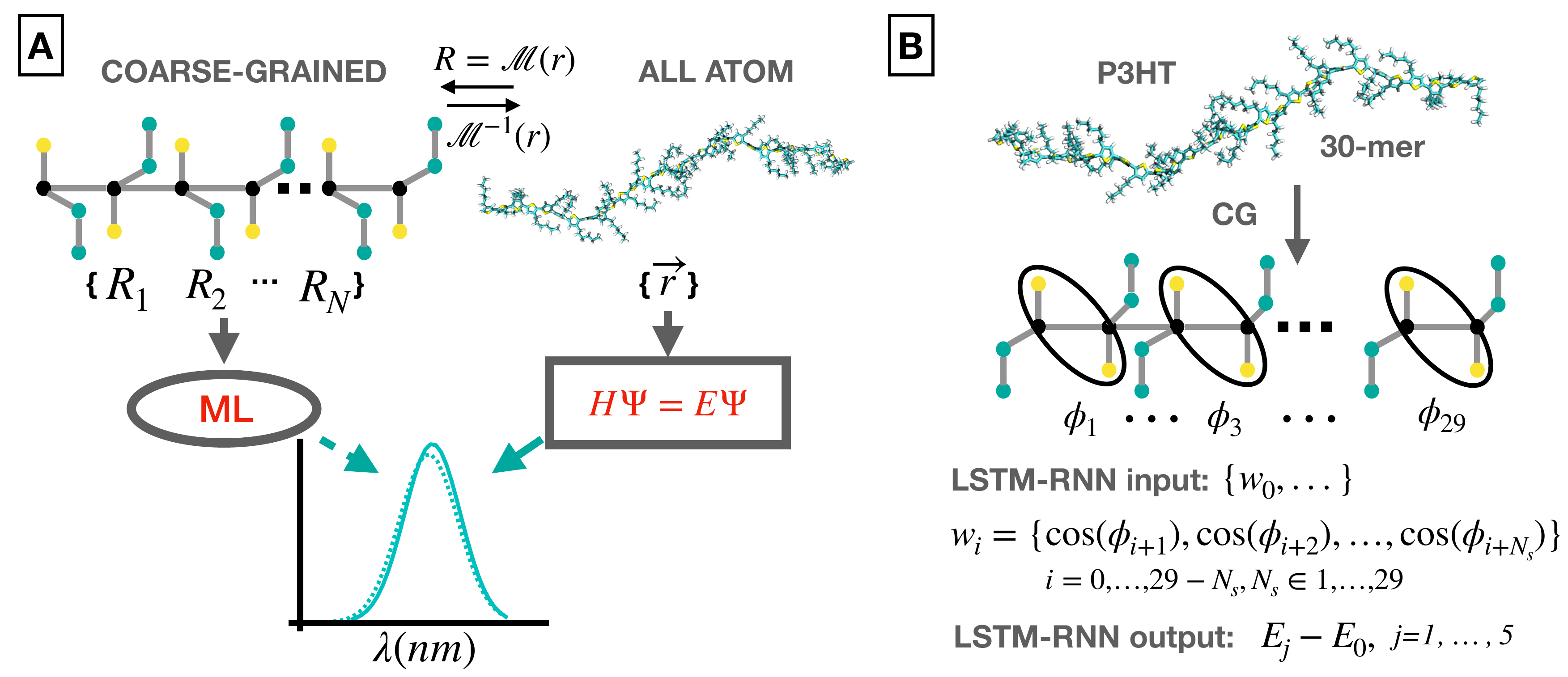}}
  \caption{A. Predicting spectra of conjugated polymers: in contrast to the traditional quantum chemistry methods which accept full atomistic description of molecules as input, we develop a machine learning (ML) method to predict spectra from incompletely known molecular structures, e.g., from coarse-grained representations. B. Given a coarse-grained model with an explicitly included term for the inter-monomer dihedrals {$\phi$} (defined in the figure by the encircled coarse-grained sites), we use an LSTM-RNN model to predict the values of excited state energies using selected sequences of $\cos(\phi)$ as inputs.}
\end{figure}

Our choice of the model was motivated by the success of a certain flavor of recurrent neural networks in capturing long-range correlations in sequential data: in the past, LSTM-RNN’s have been used successfully to capture the context, i.e., the long-range correlations in sequences of words for the purposes of language generation. Here we utilize this property of LSTM-RNN’s to relate the ‘context’ provided by the torsional conformations of poly-3-hexyl (P3HT) polymers to their absorption spectra. We train an LSTM-RNN(7) with one hidden layer and one hundred fifty (150) ‘neurons’ using a dataset which consists of input-output pairs: the 29 inter-monomer dihedrals $\phi$ and the associated value of the $j^{th}$ excited state energy relative to the ground state, i.e., $(E_j-E_0)$. The data-set was generated using molecular dynamics simulations of 50 P3HT 30-mers solvated in chlorobenzene from which about 10E6 individual structures were sampled at 10ps intervals. The spectra for the individual polymers were obtained using an established methodology using an appropriately parameterized Pariser-Parr-Pople Hamiltonian solved within Hartree-Fock and Configuration Interaction Singles approximations (8, 9). Figure 1B summarizes the structure of input-output dataset. The input consists of subsets of $\cos{\phi}$ with the inter-monomer dihedral angles $\{\phi\}$ taken from the molecular structures of our model conjugated polymer, a 30-mer of poly-3-hexylthiophene (P3HT) polymer. 

We have generate the input by including the torsions in a consecutive order using a sliding window of length $N_s$ along the backbone of the polymer and we let the training process sort out their individual importance and the importance of the correlations between them. Specifically for our model polymer, this means that the sequence of inputs consists of $29-N_s$ vectors $w_i$, where 
$w_i={\cos\phi_{i+1},\cos\phi_{i+1},…,\cos\phi_{i+N_s}}$ and the index $i$ runs over the inter-ring dihedrals one by one starting from 0 to $29-N_s$. The hyperparameter $N_s$, which is related to the localization of the wavefunction over the polymer’s backbone is determined using physical intuition and through trial and error. Interestingly, the best performance for the more localized S1 state was obtained with $N_s=6$, which matches the spread of the $S_1$ wavefunction over the backbone known from our previous work on this system (~6 monomers)(9, 10). Very small values of $N_s$ resulted in the best performance for the higher energy states (S2 and higher) whose wavefunction is delocalized over the entire 30-mer. This shows that the training processes is sensitive to the balance between the length of the input sequence ${\vec{w}}$ and the lengths of the individual $w$’s, i.e., $N_s$. The output is encoded into vectors with ones ({1}) at the positions corresponding to the values of the excitation energies on a grid.
\begin{figure}
  \centering
  \fbox{\rule[-.5cm]{0cm}{4cm} \includegraphics[width=\textwidth]{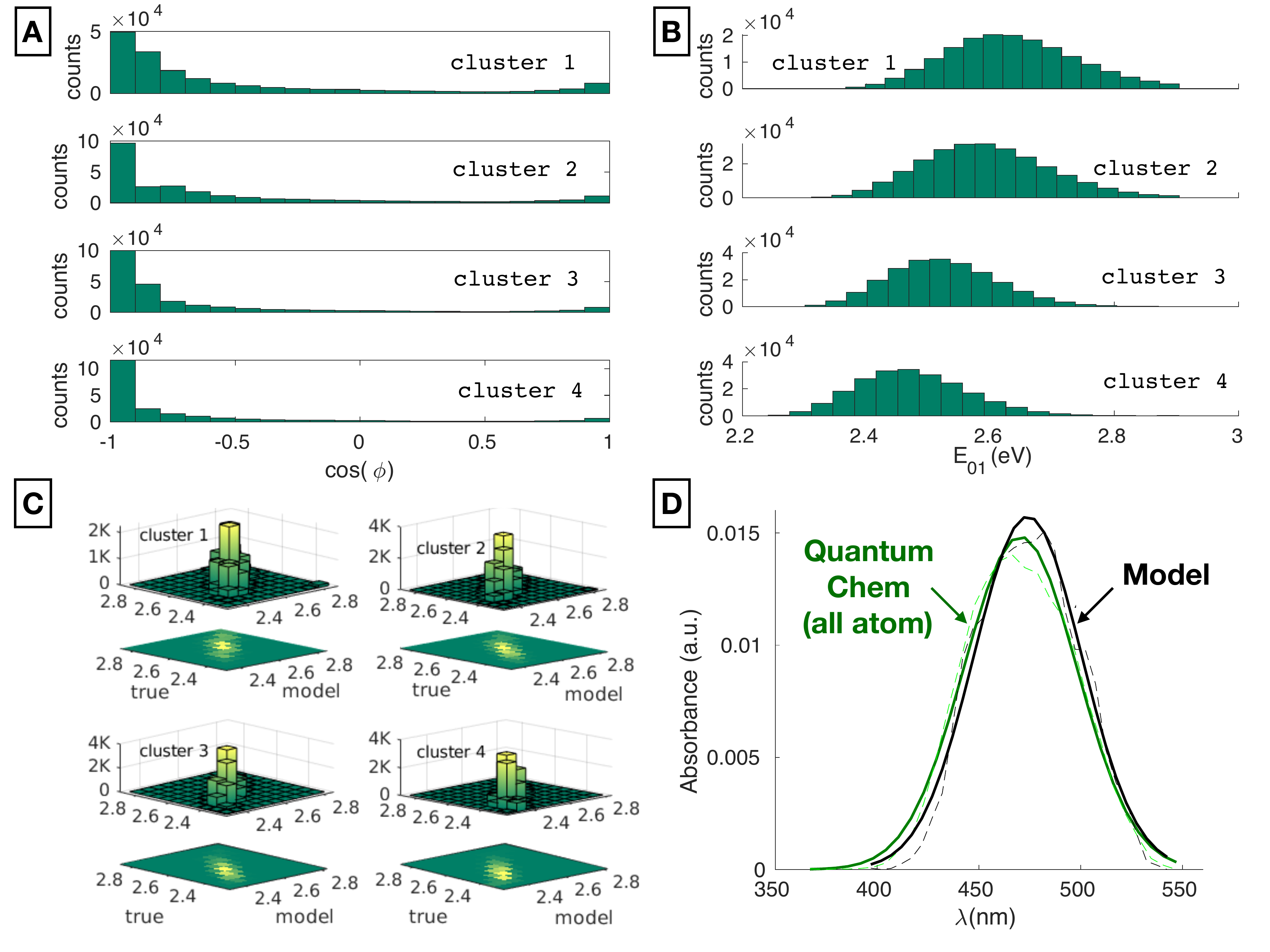}}
  \caption{A. The torsional profiles for the four distinct conformational species identified by the k-means clustering algorithm in our training data-set. B. The optical gaps, E01, in the training data-set divided into clusters shown in A. The clusters are arranged according to increasing ‘red-shift’. C. The performance of our LSTM-RNN model on a validation dataset divided into clusters shown in 2A; the Pearson coefficient is ~0.94 for all clusters. D. A comparison of spectra of our test dataset obtained using quantum chemistry from atomistic representations (green) and using the LSTM-RNN model from inter-monomer dihedrals only (black).}
\end{figure}

Figure 2A shows that k-clustering has revealed at least four distinct torsional species in the training data-set. Figure 2B shows the energies for the ground-to-first excited state transition ($E_{01}$) for each cluster. While there is a considerable overlap between the clusters, a clear red-shift of the $E_{01}$ distributions is observed indicating a considerable sensitivity of the optical gaps to the torsional profile.

For each excited state ten LSTM-RNN models had been trained and their predictions were averaged. Fig. 2C shows the correspondence between the model’s predictions and the true values of $E_{01}$ on a validation part of our training dataset. As before we are showing the correspondence broken down into clusters for increased transparency - in this case the performance on the four distinct clusters remains uniform. The Pearson coefficient was optimal for the discretized $E{0_1}$, ~$94\%$, and it deteriorated slightly for higher energy states $E_{02-05}$: $92\%-89\%$. The decrease in performance for the higher energy states is compensated by the strongly suppressed oscillator strengths associated with these states and the errors in the resulting spectra remain quite small: ~$k_BT$ at room temperature.

\begin{figure}
  \centering
  \fbox{\rule[-.5cm]{0cm}{4cm} \includegraphics[width=\textwidth]{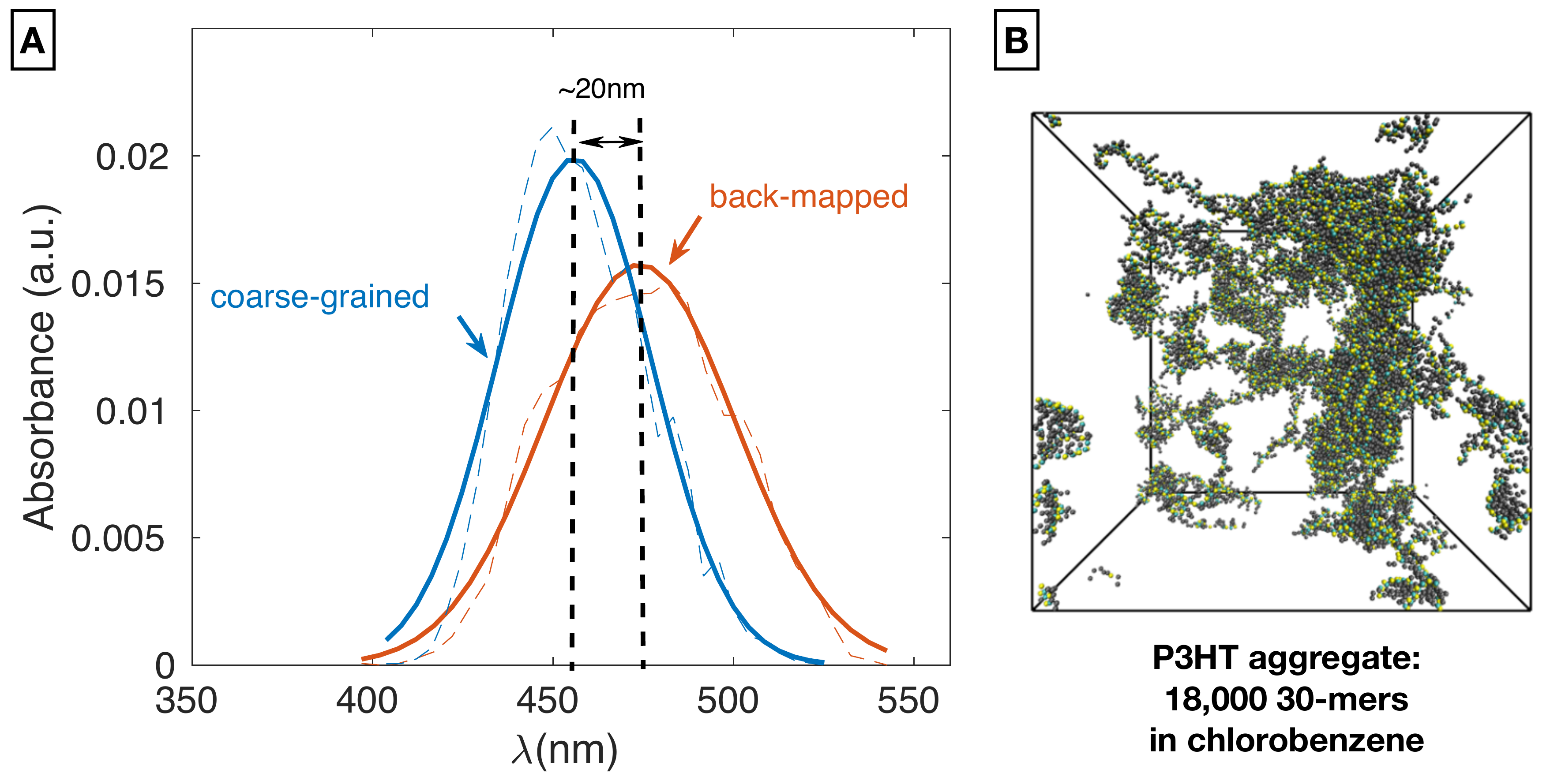}}
  \caption{A. Spectra of the test ensemble predicted by the LSTM-RNN model before (blue) and after (red) back-mapping. Note that the spectrum of the back-mapped structures is significantly red-shifted relative to the original. B. A snapshot taken from the molecular dynamics trajectory that was used to generate the test dataset, solvent not shown}
\end{figure}

We are now in a position to test the quality of our back-mapping scheme with respect to how well it preserves the absorption characteristics of the original coarse-grained structures. To this end we ran a simulation of 18000 P3HT polymers in explicit chlorobenzene (see Fig. 3B, note that the solvent is not shown). We used our model to predict the spectrum directly from the coarse-grained data using the inter-monomer dihedral angles of the polymers as input. We then back-mapped the coarse-grained structures into the atomistic representation using an ad-hoc procedure that may be considered typical for this task. 

Since we had the atomistic structures available we ran the quantum chemistry calculations on these structures as well. Fig 2D demonstrates the agreement between the predicted and the true spectra produced for our test dataset. These spectra are truncated at the five lowest energy excited states which carry the dominant fraction of the oscillator strength.

We have obtained the post-processed spectrum using the inter-monomer dihedrals from the atomistic structures. We show in Figure 3A the resulting spectra. It is clear that the post-processed ensemble (shown in red) is red-shifted relative to the original ensemble (shown in blue). The red-shift of ~20nm is similar in magnitude to shifts that are reported and physically interpreted in the spectroscopic studies of conjugated polymers. On the one hand, our approach opens the door to including quantum chemical or experimental spectroscopic information in the parameterization of coarse-grained potentials for the simulations of materials used in organic electronics or photovoltaics. On the other hand, our method can be used to optimize the back-mapping schemes. In either case the conclusions of the computational experiments which rely on coarse-grained simulations are expected to be more controlled and reliable. 
\subsubsection*{Acknowledgments}
Access to computing time on Bridges GPU-AI granted by the Extreme Science and Engineering Discovery Environment (XSEDE), which is supported by National Science Foundation grant number ACI-1548562 is gratefully acknowledged.

\section*{References}
1. 	K. Do, M. Kumar Ravva, T. Wang, J.-L. Brédas, Computational Methodologies for Developing Structure–Morphology–Performance Relationships in Organic Solar Cells: A Protocol Review. Chem. Mater. 29, 346–354 (2016).

2. 	K. N. Schwarz, T. W. Kee, D. M. Huang, Coarse-grained simulations of the solution-phase self-assembly of poly(3-hexylthiophene) nanostructures. Nanoscale. 5, 2017–2027 (2013).

3. 	R. Alessandri, J. J. Uusitalo, A. H. de Vries, R. W. A. Havenith, S. J. Marrink, Bulk Heterojunction Morphologies with Atomistic Resolution from Coarse-Grain Solvent Evaporation Simulations. J. Am. Chem. Soc. 139, 3697–3705 (2017).

4. 	S. Hochreiter, J. Schmidhuber, Long Short-Term Memory. Neural Comput. 9, 1735–1780 (1997).

5. 	I. Sutskever, O. Vinyals, Q. V Le, in Advances in Neural Information Processing Systems 27, Z. Ghahramani, M. Welling, C. Cortes, N. D. Lawrence, K. Q. Weinberger, Eds. (Curran Associates, Inc., 2014), pp. 3104–3112.

6. 	Eric Thomas Wright, thesis, University of Texas at Austin (2015).

7. 	A. Damien, TensorFlow-Examples, (available at https://github.com/aymericdamien/TensorFlow-Examples).

8. 	A. Warshel, M. Karplus, Calculation of ground and excited state potential surfaces of conjugated molecules. I. Formulation and parametrization. J. Am. Chem. Soc. 94, 5612–5625 (1972).

9. 	L. Simine, P. Rossky, Relating Chromophoric and Structural Disorder in Conjugated Polymers. J. Phys. Chem. Lett. 8, 1752–1756.

10. 	D. Raithel et al., Direct observation of backbone planarization via side-chain alignment in single bulky-substituted polythiophenes. Proc. Natl. Acad. Sci. U. S. A. 115, 2699–2704 (2018).

\end{document}